\documentclass[jap
 amsmath,amssymb,
 reprint,%
]{revtex4-1}
\usepackage{comment}

\newcommand{\bra}{\langle}
\newcommand{\ket}{\rangle}

\newcommand{\be}{\begin{equation}}
\newcommand{\ee}{\end{equation}}
\newcommand{\bea}{\begin{eqnarray}}
\newcommand{\eea}{\end{eqnarray}}
\usepackage{xcolor}
\usepackage{graphicx}
\usepackage{dcolumn}
\usepackage{bm}

\usepackage[utf8]{inputenc}
\usepackage[T1]{fontenc}
\usepackage{mathptmx}
\usepackage{etoolbox}

\makeatletter
\def\@email#1#2{%
 \endgroup
 \patchcmd{\titleblock@produce}
  {\frontmatter@RRAPformat}
  {\frontmatter@RRAPformat{\produce@RRAP{*#1\href{mailto:#2}{#2}}}\frontmatter@RRAPformat}
  {}{}
}%
\makeatother
\begin{document}

\preprint{AIP/123-QED}

\title{Lattice thermal conductivity decomposition: Peierls vs. non-Peierls contributions}
\author{Andrey Pereverzev}%
 \email{pereverzeva@missouri.edu.}
\affiliation{ 
Department of Chemistry and Materials Science $\&$  Engineering Institute, University of Missouri, Columbia, MO 65211, USA
}%

\date{\today}

\begin{abstract}
The Green–Kubo lattice thermal conductivity computed using the full classical heat current of a crystalline solid is compared with results obtained from the quadratic component of the heat current and from the commonly used Peierls heat current. In addition, thermal conductivity within the relaxation time approximation is evaluated. Three crystalline systems are investigated: solid argon, a model of solid argon with alternating masses, and  $\alpha$-quartz. For all materials considered, the thermal conductivities calculated using the quadratic and Peierls heat currents differ only slightly. In the case of $\alpha$-quartz, the optical phonon contribution to the thermal conductivity is found to exceed that of the acoustic modes. The relaxation time approximation systematically underestimates the thermal conductivity in all three systems.

\end{abstract}

\maketitle

\section{Introduction}
Calculation of the thermal conductivity (TC) tensor for  solids can
be accomplished using the Green-Kubo (GK) formalism implemented within classical
molecular dynamics.\cite{Kubo} The TC tensor in the GK framework is given by
\be
\kappa^{\alpha\beta}=\frac{1}{k_BT^2V}\int_0^{\infty}\bra j^{\alpha}(t)j^{\beta}(0)\ket dt, \label{GK}
\ee
where $j^{\alpha}$ denotes $\alpha$th component of one of the heat current, $V$ and $T$ are system volume and temperature, and brackets indicate averaging over an equilibrium ensemble. 
The integrand in Eq.
(\ref{GK}) defines the tensoral time-dependent heat current correlation function (HCCF). The HCCF shows decay as a function of time which ensures convergence of the integral in Eq. (\ref{GK}).  The specific functional form of this decay depends on each particular solid. For monatomic crystals\cite{Ladd,Che,McGaughey1} the HCCF can usually be approximated by a superposition of monotonically decaying exponentials. For more complex polyatomic crystals with unit cells containing tens
or hundreds of atoms the exponential terms in the HCCF are superposed
with decaying oscillatory terms with various frequencies.\cite{McGaughey2,Izvekov,PereverzevSewell2022} This  complex behavior of the HCCF can be qualitatively explained using detailed analysis of the heat current expression whereby the current is expanded in a Taylor series of atomic displacements and velocities.\cite{PereverzevSewell2018,Baroni,PereverzevSewell2022} It was shown theoretically and verified numerically that the linear terms of a Taylor expansion do not contribute to TC but partially account for  oscillations of the HCCF. \cite{Baroni,PereverzevSewell2022} Additionally, oscillation of the HCCF can arise form quadratic terms of a Taylor series expansion of the heat current: when converted into the normal mode space  the quadratic terms can be separated into the so-called Peierls terms and the non-Peierls contributions and it is the latter  that lead to oscillatory terms in the HCCF.\cite{PereverzevSewell2018,Simoncelli}

Heat current decomposition into various terms in the phonon picture and the corresponding TC contributions were considered by several groups. Sun and Allen \cite{SunAllen} studied TC in two-dimensional triangular lattice calculated using different components of the total heat current. Simoncelli et al.\cite{Simoncelli} studied
TC decomposition in the perovskite CsPbBr$_3$. 

Here we study dependence of the numerically calculated TC on the choice of the heat current, viz. the full heat current and its approximations for three crystalline systems with the goal of understanding how TC depends on the choice of heat current and system complexity.

\section{Heat current decomposition}
The heat current for a system consisting of  $N$ atoms is defined by the following expression:
\be
{\bf J}=\sum_{i=1}^N\left(\epsilon_i{\bf v}_i-{\bf S}_i\cdot{\bf v}_i\right), \label{curdef}
\ee
where $\epsilon_i$  is the energy of atom $i$ and ${\bf S}_i$ is the per-atom stress tensor of that atom. \cite{Plimpton} 
The tensor ${\bf S}_i$ in Eq. (\ref{curdef}) multiplies
atomic velocity ${\bf v}_i$ as a $3\times3$ matrix multiplies a vector, to yield a vector. The atomic energy $\epsilon_i$ is given by  
\begin{equation}
\epsilon_i=\frac{m_i |{\bf v}_i |^2}{2}+u_i, \label{locen}
\end{equation}
where the first and second terms on the right-hand side are, respectively, the atomic kinetic and potential energies.
There is a well-known ambiguity in defining $u_i$ for any system with interatomic potential terms that arise from different ways of partitioning potential energy among different atoms. \cite{Hardy, Allen,Schelling} It was shown theoretically \cite{Marcolongo} and numerically \cite{Schelling,Marcolongo} that this ambiguity does not affect TC values obtained using GK approach. 
Here we use the definition of $u_i$ used in LAMMPS. \cite{Plimpton}

  For further analysis the expression for the full  current  (\ref{curdef})
is expanded in a Taylor series of atomic velocities and atomic displacements from equilibrium positions to the second order and written here for $\alpha$th  Cartesian component
\bea
 J^{\alpha}&=&\sum_{i=1}^N u_i^0{v}_i^{\alpha}-\sum_{i=1}^N\sum_{\beta} \left({S}_i^{\alpha\beta}\right)^0v_i^{\beta}   \nonumber \\
& &+ \sum_{i=1}^N\sum_{j=1}^N\sum_{\beta} {\left(\frac{\partial u_i}{\partial x_j^{\beta}}\right)}^0\left(x_j^{\beta}-x_j^{\beta,0}\right){v}_i^{\alpha}\nonumber \\
& &-\sum_{i=1}^N\sum_{j=1}^N\sum_{\beta,\gamma} \left(\frac{\partial{S}_i^{\alpha\beta}}{\partial x_j^{\gamma}}\right)^0\left(x_j^{\gamma}-x_j^{\gamma,0}\right)v_i^{\beta} + ... \;\;.\label{expand}
\eea
Here the superscript $0$ means that the corresponding quantities are evaluated at the equilibrium values of atomic coordinates, $x_j^{{\alpha},0}$. The first two terms on the right-hand side of Eq. (\ref{expand}) are linear in atomic velocities and do not depend on atomic displacements. These terms are usually non-vanishing in polyatomic crystals \cite{Pereverzev2020} and can lead to strong oscillatory contributions to the heat current correlation function. However, it can be shown theoretically and numerically that the integrals over these oscillatory contributions vanish \cite{Marcolongo,Pereverzev2020}. Thus, the first two terms on the right-hand side of Eq. (\ref{expand}) do not contribute to TC and can be dropped to smooth out the heat current correlation functions. 

The third and fourth terms on the right-hand side of Eq. (\ref{expand}) represent the quadratic part of the heat current, ${\bf J}_{\text{quad}}$. 
Peierls heat current constitute a part of the quadratic current. It is obtained by converting the stress-dependent part of the quadratic current (given by the fourth term on the right-hand side of Eq. (\ref{curdef})) to normal mode coordinates and keeping only the terms that are diagonal in the phonon branch index as discussed in detail in Ref.  \cite{PereverzevSewell2018}. The resulting expression  for Peierls heat current is 
\be
J^{\alpha}_{\text{Peierls}}=\sum_{{\bf k},s}v^{\alpha}({\bf k},s)E({\bf k},s). \label{Peierls}
\ee
Here $E({\bf k},s)$ is the harmonic energy of a mode specified with the wave vector ${\bf k}$ and branch $s$  and $v^{\alpha}({\bf k},s)$ is $\alpha$th Cartesian component of the group velocity of that mode.
Although Eq. (\ref{Peierls}) represents only a part of the total heat current (\ref{curdef}), it is often used in theoretical analysis of the heat transport  in the normal mode picture. 

In this work we compare thermal conductivities for a few crystalline materials calculated using three forms of the heat current: the full heat current given by Eq. (\ref{curdef}), the quadratic heat current given by the quadratic terms in Eq. (\ref{expand}), and the Peierls heat current (\ref{Peierls}). In the latter case both the full
Peierls heat current as well as its parts due to acoustic and
optical branches were considered.
For all three forms of the current the TC tensor is calculated using the Green-Kubo approach based on Eq. (\ref{GK}).

In addition to calculating TC using Eq. (\ref{GK}) for three forms of the heat current we also calculated TC in the relaxation time approximation (RTA), whereby the GK approach is applied to fluctuations of the individual normal-mode heat currents, namely,  
 \bea 
& &\kappa^{\alpha \beta}_{\text{RTA}}=\frac{1}{k_BT^2V}\sum_{{\bf{k}},s}\Big[\int_0^\infty  v^{\alpha}({\bf k},s)v^{\beta}({\bf k},s)\nonumber \\
& &\times\big(E({\bf k},s,t)-\overline{E}({\bf k},s)\big)  \big(E({\bf k},s,0)-\overline{E}({\bf k},s)\big) dt\Big]. \label{rwa}
 \eea
Here $E({\bf k},s,t)$ denotes the harmonic mode energy of mode with wave vector $\bf k$ and branch $s$ at time $t$ and  $\overline{E}({\bf k},s)$ is the ensemble-averaged value of the same quantity. The ensemble-averaged classical harmonic energy of a normal mode is expected to be exactly $k_BT$. However, actual time-averaged  harmonic energy values for individual modes (which differed very slightly from $k_BT$) were used in  Eq. (\ref{rwa}).

\section{Models and simulation details}\label{sec3}
We applied our analysis to three crystalline systems. The first system considered was solid argon (SA) modeled with the Lenard-Jones potential,
\be
U_{{\text{Ar}}}(r_{ij})=4\epsilon\left[\left(\frac{\sigma}{r_{ij}}\right)^{12}-\left(\frac{\sigma}{r_{ij}}\right)^{6}\right], \label{LJ}
\ee
where $r_{ij}$ is the distance between atoms $i$ and $j$ and $\sigma$ and $\epsilon$ are parameters specifying the potential.  The following parameters were used: $\sigma=3.4$ {\AA} and $\epsilon=1.67 \times 10^{-23}$ J \cite{Ashcroft,McGaughey1}. The atomic  mass of argon atom was taken to be 39.948. The SA forms face centered cubic (fcc) lattice whose phonon spectrum consists of three acoustic branches \cite{Ashcroft}.

The second system studied here was closely related to SA but differed  from it by the presence of both acoustic and optical branches in the phonon spectrum. The interaction potential for the second model was identical to the SA system (Eq. \ref{LJ}) with the same potential parameters.  In order to  achieve splitting of the phonon spectrum into acoustic and optical branches, two alternating atomic masses which differ by a factor of 10  were used, namely
 7.26327 and 72.6327. The average of these two values is 39.948, which corresponds the atomic mass of argon. We refer to this crystal as the solid argon with alternating masses (SAAM) model. Even though the atomic positions in the SAAM model are the same as in the fcc SA model, the SAAM crystal  belongs to the P4/mmm space group which corresponds to a simple tetragonal lattice with two atoms per unit cell. To make the comparison of thermal conductivities of the fcc solid argon and the tetragonal SAAM crystals as direct as possible we treat the solid argon as a tetragonal lattice with two atoms per unit cell. Figure 1 shows unit cells of solid argon and the SAAM model. The two unit cells have identical lattice parameters because the SA and SAAM models have identical interaction potentials and average kinetic energies. Because of the identical lattice parameters and average atomic masses, the SA and SAAM crystals have identical densities.
 
 Figure \ref{Figure2} shows phonon spectra for the SA and SAAM models. Treating the SA model as a tetragonal lattice effectively folds the Brillouin zone of the fcc lattice (given by a truncated octahedron) into a tetragonal  Brillouin zone of the tetragonal lattice with the higher frequency acoustic modes of solid argon becoming three optical-like branches in the tetragonal Brillouin zone as shown in red in Fig. \ref{Figure2}. We call the three low and the three high frequency branches of the SA model (treated as a tetragonal lattice) as the low and high frequency branches, respectively, in the analysis below.
The simulation cell sizes for both SA and SAAM crystals were chosen to be $7\times7\times5$ unit cells shown in Fig. 1. 
The simulations were performed at $T=20$ K and zero pressure. For the SA model the analysis was based on 10 60 ns NVE trajectories. In the case of the SAAM model 40 60 ns NVE trajectories were used. The larger number of trajectories for the SAAM model was needed because GK approach generally shows slower convergence in polyatomic crystals compared to  monatomic ones.  The simulation time step was set to 2 fs; data was dumped every 30 fs for both models. 
\begin{figure}
 \includegraphics[width=\columnwidth]{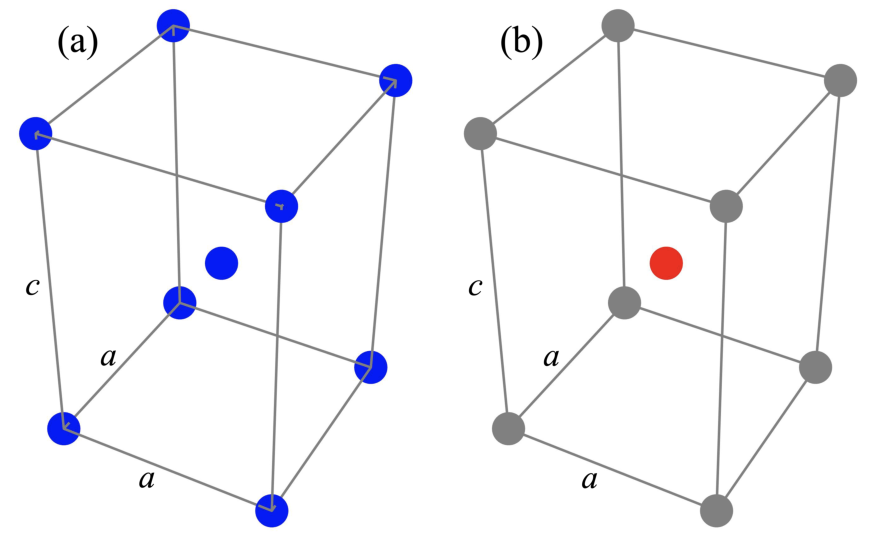}  
 \caption{\label{Figure1} (a) Solid argon treated as a tetragonal lattice. The tetragonal unit cell with two atoms per unit cell is shown; $a=3.7465 $ \AA, $c=\sqrt{2}a$. The lattice parameters correspond to $T=20$ K and zero pressure. (b) The  tetragonal unit cell of the SAAM model with heavy and light atoms shown in gray and red, respectively. The unit cell parameters are the same as in (a).
}
 \end{figure}

\begin{figure}
 \includegraphics[width=\columnwidth]{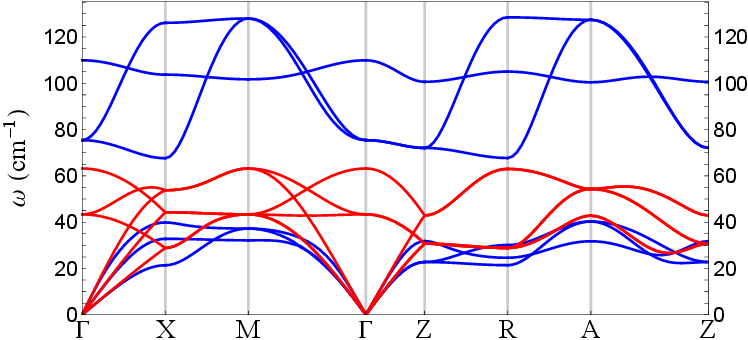}  
 \caption{\label{Figure2} Phonon branches of the SA (red) and SAAM (blue) models for selected Brillouin zone directions of the tetragonal lattice. Note the frequency gap between the acoustic and  optical branches of the SAAM model.
}
 \end{figure}

\begin{figure}
 \includegraphics[width=\columnwidth]{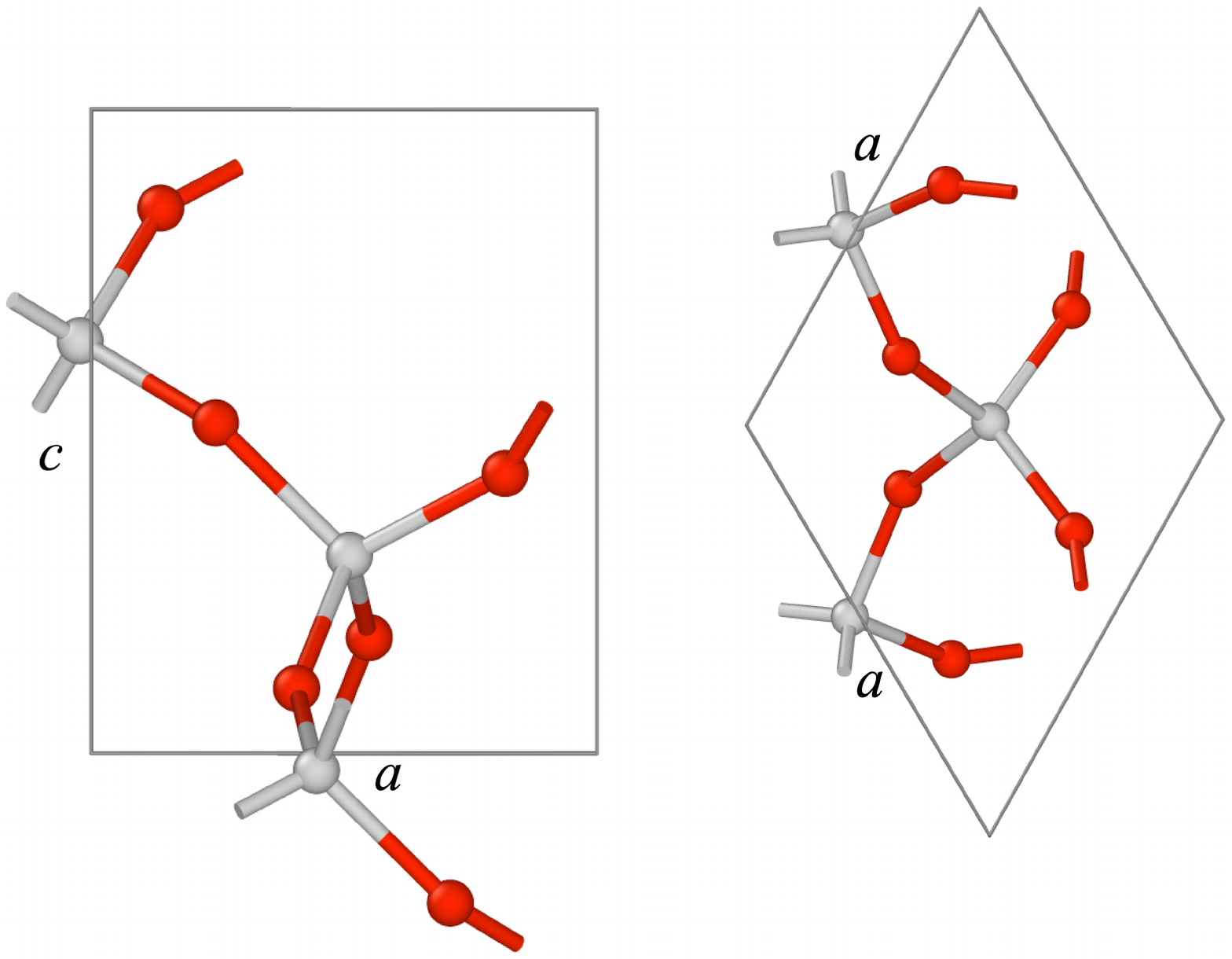}  
 \caption{\label{Figure3} The unit cell of $\alpha$-quartz. Left: viewed along {\it{a}}. Right: viewed along {\it{c}}. Si and O atoms are shown in gray and red, respectively. Lattice parameters $a=4.96264$ {\AA}  and $c=5.4652$ {\AA}.
}
 \end{figure}

The third system studied was $\alpha$-quartz, a trigonal crystal with nine atoms per unit cell and 27 phonon branches. The unit cell of  $\alpha$-quartz is shown in Fig. \ref{Figure3}.

$\alpha$-Quartz has been  studied both experimentally \cite{Touloukian},  and computationally \cite{McGaughey2},  with MD results showing good agreement with experiment. It is known that the $\alpha$-quartz heat current correlation function  exhibits strong oscillations. It was argued that these oscillations contribute significantly to the thermal conductivity.\cite{McGaughey2} $\alpha$-Quartz was modeled with the pairwise Beest–Kramer–van Santen (BKS) interatomic potential \cite{BKS,McGaughey2}, in which the potential energy  between atoms $i$
and $j$ is given by
\be
U_{\text{quartz}}(r_{ij})=\frac{q_iq_j}{r_{ij}} + A_{ij}\exp(-b_{ij}r_{ij})-\frac{c_{ij}}{r_{ij}^6}, \label{BKS}
\ee
where $q_i$ is an atomic charge, $A_{ij}$, $b_{ij}$, and $c_{ij}$ are constants
specified by the types of atoms $i$ and $j$ (either oxygen or
silicon), and $r_{ij}$ is the distance between atoms $i$ and $j$. The values of parameters in (\ref{BKS}) are listed in Ref. \cite{BKS}.
Crystal cell comprising $5\times5\times5$ unit cells (1125 atoms) was studied to  keep the system computationally manageable. Similar system sizes were used by others.\cite{McGaughey2}
Simulations of $\alpha$-quartz were performed at $T=300$ K and zero pressure.
TC decomposition for $\alpha$-quartz was based on 40 statistically independent 0.5 ns NVE trajectories which were computed using 0.5 fs timestep. The data was dumped every 2 fs.

Molecular dynamics simulations for all three models were performed using the LAMMPS package.\cite{Plimpton} 

The harmonic heat current as a function of time was calculated from the quadratic terms of Eq. (\ref{expand}).
The derivatives of local energy and per-atom stress at the potential energy minimum were obtained numerically using finite differences by applying finite atomic displacements to the minimum energy crystal geometry.

The Peierls heat current  (\ref{Peierls}) was obtained by calculating harmonic energies of individual normal modes as functions of time along with the corresponding group velocities. More specifically, the normal mode analysis was performed for each system using minimum energy geometry to obtain normal modes and mode frequencies. 
Dumped Cartesian atomic displacements and velocities were converted to normal mode coordinates and momenta from which  harmonic mode energies were calculated. \cite{PereverzevSewell2018} Group velocities are usually defined as derivatives of mode frequencies with respect to wave vector, $v_s^{\alpha}(\bf{k})=\partial \omega_s(\bf{k})/\partial k^{\alpha}$. Since the systems used in simulations were finite, wave vectors were discrete variables and partial derivatives were not well defined. To this end, a definition of the group velocity applicable to finite systems in which it is expressed  in terms of eigenstates of the dynamical matrix and the dynamical matrix itself was used.\cite{PereverzevSewell2018,Allen,Hardy}

\section{Results and discussion}
 The main results are summarized in Tables \ref{table1}, \ref{table2}, and \ref{table3}.
\begin{table}
\caption{\label{table1}Thermal conductivities of the SA model at 20 K and zero pressure. Units are W m$^{-1}$K$^{-1}$.}
\begin{ruledtabular}
\begin{tabular}{ l c}

$\kappa_{\text{full}}$ & $1.35\pm 0.06$ \\
$\kappa_{\text{quad}}$ & $1.19\pm 0.03$ \\
$\kappa_{\text{Peierls}}$ & $1.19\pm 0.03$ \\
$\kappa_{\text{Peierls,low}}$ & $0.69\pm 0.02$ \\
$\kappa_{\text{Peierls,high}}$ & $0.37\pm 0.01$ \\
$\kappa_{\text{Peierls,low}}+\kappa_{\text{Peierls,high}}$ & $1.05\pm 0.03$ \\
$\kappa_{\text{RTA}}$ & $1.01\pm 0.02$ \\
$\kappa_{\text{RTA,low}}$ & $0.67\pm 0.01$ \\
$\kappa_{\text{RTA,high}}$ & $0.35\pm 0.01$ \\
\end{tabular}
\end{ruledtabular}
\end{table}
Table \ref{table1} lists thermal conductivities of the SA system. The TC of the fcc lattice is isotropic; thus, the conductivities for different currents shown in Table \ref{table1} are direction-independent. The TC calculated using the full heat current has the highest value. The TC calculated from the quadratic  and Peierls currents are the same and slightly lower than $\kappa_{\text{full}}$. $\kappa_{\text{Peierls,low}}$ and $\kappa_{\text{Peierls,high}}$ are the thermal conductivities obtained using low and high frequency modes as defined in Sec. \ref{sec3}. As can be expected, $\kappa_{\text{Peierls,low}}$ is higher than $\kappa_{\text{Peierls,high}}$ because typical group velocities are higher at low frequencies. Note that $\kappa_{\text{Peierls,low}}+\kappa_{\text{Peierls,high}}$ is 
slightly lower than $\kappa_{\text{Peierls}}$, the difference arises from the cross-correlation terms between the low and high frequency mode contributions in the GK formula. The RTA approximation gives values that are approximately 15\% lower than  Peierls conductivity. The RTA conductivities for low and high frequency modes are also slightly lower than the corresponding Peierls conductivities. 

The SAAM model thermal conductivities are shown in Table \ref{table2}. Since the SAAM crystal is tetragonal TC is anisotropic. Conductivities along {\textbf{\textit{a}}} are substantially higher than the ones along {\textbf{\textit{c}}}. For both directions the conductivities for quadratic current is lower than the full current ones. Peierls conductivities are very slightly lower than the quadratic ones. Acoustic  Peierls conductivity is approximately three and a half times higher than the optical one for the {\textbf{\textit{a}}} direction; for the {\textbf{\textit{c}}} direction the optical Peierls conductivity is vanishingly small. The RTA underestimates all other forms of TC for both directions. In comparison to the isotropic SA model, the SAAM TC is lower both for {\textbf{\textit{a}}} and {\textbf{\textit{c}}} directions.

\begin{table}
\caption{\label{table2}Thermal conductivities of the SAAM model at 20 K and zero pressure. Units are W m$^{-1}$K$^{-1}$.}
\begin{ruledtabular}
\begin{tabular}{ l c}
$\kappa_{\text{full}}^{a}$ & $1.26\pm 0.02$ \\
$\kappa_{\text{quad}}^{a}$ & $1.10\pm 0.02$ \\
$\kappa_{\text{Peierls}}^{a}$ & $1.07\pm 0.02$ \\
$\kappa_{\text{Peierls,ac}}^{a}$ & $0.82\pm 0.02$ \\
$\kappa_{\text{Peierls,opt}}^{a}$ & $0.231\pm 0.005$ \\
$\kappa_{\text{Peierls,ac}}^{a}+\kappa_{\text{Peierls,opt}}^{a}$ & $1.05\pm 0.02$ \\
$\kappa_{\text{RTA}}^{a}$ & $0.939\pm 0.006$ \\
$\kappa_{\text{RTA,ac}}^{a}$ & $0.710\pm 0.005$ \\
$\kappa_{\text{RTA,opt}}^{a}$ & $0.229\pm 0.001$ \\
\hline
$\kappa_{\text{full}}^{c}$ &$0.82\pm 0.03$  \\
$\kappa_{\text{quad}}^{c}$ &$0.74\pm 0.02$  \\
$\kappa_{\text{Peierls}}^{c}$ &$0.73\pm 0.02$  \\
$\kappa_{\text{Peierls,ac}}^{c}$ &$0.73\pm 0.02$  \\
$\kappa_{\text{Peierls,op}}^{c}$ &$0.0022\pm 0.0001$  \\
$\kappa_{\text{Peierls,ac}}^{c}+\kappa_{\text{Peierls,opt}}^{c}$ &$0.73\pm 0.02$  \\
$\kappa_{\text{RTA}}^{c}$ & $0.71\pm 0.02$ \\
$\kappa_{\text{RTA,ac}}^{c}$ & $0.71\pm 0.02$ \\
$\kappa_{\text{RTA,opt}}^{c}$ & $0.0024\pm 0.0001$ \\
\end{tabular}
\end{ruledtabular}
\end{table}

Table \ref{table3} shows TCs of $\alpha$-quartz along {\textbf{\textit{a}}} and {\textbf{\textit{c}}}.
The quadratic TC along {\textbf{\textit{a}}} is slightly lower than the full TC of 9.3 W m$^{-1}$K$^{-1}$; the Peierls TC is higher
that the quadratic one but very slightly lower that the full TC. Surprisingly, optical contributions to Peierls TC along {\textbf{\textit{a}}} are more than three times higher than the acoustic ones.  There is some correlation between acoustic and optical modes for this direction since $\kappa_{\text{Peierls,ac}}^{a}+\kappa_{\text{Peierls,opt}}^{a}<\kappa_{\text{Peierls}}^{a}$. The RTA approximation for TC long {\textbf{\textit{a}}} gives values similar to the Peierls TC.

The quadratic TC along {\textbf{\textit{c}}} is slightly higher than the full TC of 13.1 W m$^{-1}$K$^{-1}$; the Peierls TC is lower
that the quadratic one but very slightly higher that the full TC. Note, however, that the differences between the three values are within numerical uncertainty. The optical modes Peierls TC is slightly higher than the acoustic modes one for this direction. There is virtually no correlation between acoustic and optical modes. The RTA approximation for TC values long {\textbf{\textit{c}}} are similar to the ones for the Peierls TC.

The full conductivities along {\textbf{\textit{a}}} and {\textbf{\textit{c}}} reported in Table \ref{table3} are  somewhat higher than the GK MD-based results reported in Ref. \cite{McGaughey2} for 300 K and zero pressure and the same force field. The difference possibly stems from the effective truncation of the  Coulombic potential as well as the smaller simulation cell size used in that reference. 

Another important distinction from the results of Ref. \cite{McGaughey2} is the  values of estimated contributions to TC from optical and  acoustic modes: as discussed above, depending on direction, the optical mode contribution to Peierls TC found in this work is significant or dominant, whereas
the optical modes TC  reported in Ref. \cite{McGaughey2} represents a small fraction of the total TC.
Ref. \cite{McGaughey2} used a heuristic 
approach based on the heat current correlation function fitting to an ad hoc function that includes optical and acoustic contributions; a similar approach was used earlier  by Che et al. to partition diamond TC. \cite{Che} By contrast,  we use direct evaluation of the optical and acoustic mode contributions to the Peierls TC without using any additional assumptions or fittings. The difference between the results of Ref.\cite{McGaughey2} and this work shows  limitations of the fitting approaches based on the ad hoc expressions for the heat current correlation functions:
 as discussed
above, a significant part of the oscillations in the correlation functions does not contribute to TC because
the time integral over them vanishes; therefore, fitting oscillatory
functions to the HCCF data may lead to incorrect estimates of the
thermal conductivity because doing so involves fitting functions
with finite time integrals to the data, some parts of which have
vanishing time integrals.\cite{PereverzevSewell}
\begin{table}
\caption{\label{table3}Thermal conductivities of $\alpha$-quartz at 300 K and zero pressure. Units are W m$^{-1}$K$^{-1}$. }
\begin{ruledtabular}
\begin{tabular}{ l c}
$\kappa_{\text{full}}^{a}$ & $9.3\pm 0.5$ \\
$\kappa_{\text{quad}}^{a}$ & $9.0\pm 0.5$ \\
$\kappa_{\text{Peierls}}^{a}$ & $9.2\pm 0.5$ \\
$\kappa_{\text{Peierls,ac}}^{a}$ & $2.1\pm 0.1$ \\
$\kappa_{\text{Peierls,opt}}^{a}$ & $6.7\pm 0.4$ \\
$\kappa_{\text{Peierls,ac}}^{a}+\kappa_{\text{Peierls,opt}}^{a}$ & $8.8\pm 0.5$ \\
$\kappa_{\text{RTA}}^{a}$ & $9.2\pm 0.2$ \\
$\kappa_{\text{RTA,ac}}^{a}$ & $2.4\pm 0.1$ \\
$\kappa_{\text{RTA,opt}}^{a}$ & $6.8\pm 0.1$ \\
\hline
$\kappa_{\text{full}}^{c}$ &$13.1\pm 0.9$  \\
$\kappa_{\text{quad}}^{c}$ &$13.4\pm 0.9$  \\
$\kappa_{\text{Peierls}}^{c}$ &$13.2\pm 0.9$  \\
$\kappa_{\text{Peierls,ac}}^{c}$ &$6.3\pm 0.4$  \\
$\kappa_{\text{Peierls,op}}^{c}$ &$6.8\pm 0.5$  \\
$\kappa_{\text{Peierls,ac}}^{c}+\kappa_{\text{Peierls,opt}}^{c}$ &$13.1\pm 0.9$  \\
$\kappa_{\text{RTA}}^{c}$ & $13.0\pm 0.2$ \\
$\kappa_{\text{RTA,ac}}^{c}$ & $6.3\pm 0.1$ \\
$\kappa_{\text{RTA,opt}}^{c}$ & $6.7\pm 0.1$ \\
\end{tabular}
\end{ruledtabular}
\end{table}

Comparing the results for SA/SAAM models with those for $\alpha$-quartz
we note that both SA and SAAM models  show a 10-13\% lower value for the quadratic and Peierls TC compared to the full TC, but there is a much smaller difference in the case of $\alpha$-quartz.
These differences most likely arise from contributions to TC from anharmonic (i.e., cubic and higher-order) terms in the heat current. Such contributions are not explicitly analyzed in this work. The larger difference observed for the SA and SAAM models, compared to the smaller one for $\alpha$-quartz, may originate from differences in the forms of their respective Hamiltonians, which in turn lead to distinct anharmonic heat-current contributions to TC. It is also worth noting that the simple measure of anharmonicity discussed below  may be too crude to capture these effects, as anharmonic heat-current contributions can depend sensitively on the specific phonon spectrum and on various resonance conditions, analogous to the situation encountered in the phonon Boltzmann equation.\cite{Ziman}

For both directions in the SAAM model, the optical contribution to the Peierls conductivity is much smaller than the acoustic one, while the opposite is observed for $alpha$-quartz.  The most likely reason for this difference is the larger number of optical branches in $\alpha$-quartz (24 optical branches) compared to the SAAM model (3 optical branches). This makes it possible for more heat to be carried by optical modes in $\alpha$-quartz in comparison to the SAAM model.

Since a finite TC in solids is caused by anharmonic couplings among normal modes \cite{Ziman}, it is instructive to estimate magnitude of crystal anharmonicity for the temperature and pressure values used in the simulations.
As a measure of anharmonicity we use 
\be 
\eta=\frac{\Big(\overline{U}_{{\text{pot,adj}}}-(N-3)k_BT/2\Big)}{(N-3)k_BT/2},
\ee
the relative difference between the adjusted average full potential energy 
$\overline{U}_{{\text{pot,adj}}}$ and the harmonic potential energy. Adjusted here means that the potential energy at the potential energy minimum is subtracted from the full potential energy so that  ${U}_{{\text{pot,adj}}}=0$ at $T=0$. In a purely harmonic crystal with periodic boundaries the average potential energy is equal to $(N-3)k_BT/2$, where $N$ is the number of atoms. The value of $\eta$ for SA and SAAM models at 20 K and zero pressure was approximately $0.01$ or $1\%$ whereas $\eta$ value for $\alpha$-quartz was approximately $0.016$ or $1.6\%$. Thus, while the values of  $\eta$ are small, the crystal anharmonicity is sufficient to cause finite and sizable TC for all three systems studied here.

\section{Conclusions}
We studied contributions of different forms of the heat current to TC using the GK approach. For the models studied the TC for the full heat current does not differ significantly from the ones for the quadratic and Peierls heat currents. Note that all three systems studied have units cells composed of a few atoms. Our more recent results\cite{Pereverzev2026}  show that for systems with large unit cells, namely $\beta$-HMX (with 56 atoms per unit cell) and $\alpha$-RDX (with 168 atoms per unit cell)  Peierls  TC is substantially smaller than the full current TC. Similar trend was observed in Ref. \cite{Simoncelli} where TC of the perovskite CsPbBr$_3$ was studied. 

\begin{acknowledgments}
This research was funded by Air Force Office of Scientific Research, Grant No. FA9550-22-1-0212. The author is grateful to Tommy Sewell for useful discussions and Steven Kelley for providing SAAM model crystal structure.
\end{acknowledgments}

\section*{Data Availability Statement}

The data that support the findings of this study are available from the corresponding author upon reasonable request.


%

\end{document}